\documentclass[lettersize,journal]{IEEEtran}
\usepackage{amsmath,amsfonts}
\usepackage{algorithmic}
\usepackage{algorithm}
\usepackage{array}
\usepackage[caption=false,font=normalsize,labelfont=sf,textfont=sf]{subfig}
\usepackage{textcomp}
\usepackage{stfloats}
\usepackage{url}
\usepackage{verbatim}
\usepackage{graphicx}
\usepackage{cite}
\usepackage[caption=false]{subfig}
\begin{document}

\title{Spatially-Adaptive Neural-Network Based Extraction of Significant Wave Heights from HF Radar Data}

\author{Senal Chandrasekara and Reza Shahidi%
\thanks{This work was supported by the Natural Sciences and
Engineering Research Council of Canada (NSERC) through Discovery
Grant RGPIN-2020-07155 awarded to Reza Shahidi.
(Corresponding author: Senal Chandrasekara.)}%
\thanks{Senal Chandrasekara and Reza Shahidi are with the
Department of Electrical and Computer Engineering,
Memorial University of Newfoundland, St. John's, NL, Canada
(e-mail: cchandraseka@mun.ca; rshahidi@mun.ca).}}
\maketitle

\begin{abstract}
Significant wave height (SWH) retrieval from high-frequency
(HF) radar data conventionally relies on Doppler-spectrum
processing and interpretation of the weak second-order
scattering continuum. This article investigates adaptive
spatial processing for time-domain SWH estimation using
ordered envelope statistics. In our previous study, a physics-guided linear
ordered-statistics estimator (LOSE), motivated by
first-order rough-surface scattering assumptions and
radar-envelope statistics, mapped the largest rank-ordered
magnitudes of Bartlett-beamformed voltages to buoy-observed
SWH through least-squares regression. The LOSE framework is
extended using three trainable complex spatial filters
initialized around the buoy look direction. Ordered
magnitudes are extracted independently from each spatial
output, concatenated, and mapped to SWH through
least-squares regression, while the spatial weights are
optimized using gradient-based updates. Both estimators
use causal sliding-window averaging. Evaluation used
measurements from a 13.385-MHz, 12-receiver element WERA radar
at Argentia, Canada, paired with
directional-wave buoy observations. Under a common
chronological training--evaluation partition, the
selected adaptive configuration reduced
root-mean-square error by approximately
$14.3\%$, while maintaining the Pearson correlation
coefficient at $0.949$. These results
support the potential of adaptive multi-beam processing
to improve time-domain SWH estimation relative to
fixed Bartlett-based LOSE. Feature counts and smoothing
windows were selected using the evaluation interval;
independent testing is therefore needed to establish
the generality of the improvement.
\end{abstract}


\section{Introduction}
\label{sec:introduction}

Significant wave height (SWH) is an important measure
of sea-state severity for maritime navigation, coastal
engineering, and offshore operations. Wave buoys provide
local reference measurements, but their spatial coverage
is limited. Shore-based high-frequency (HF) radar offers
the potential to monitor ocean conditions over extended
coastal areas by measuring electromagnetic backscatter
from the sea surface.

\begin{figure}[!t]
    \centering
    \includegraphics[width=0.92\columnwidth]{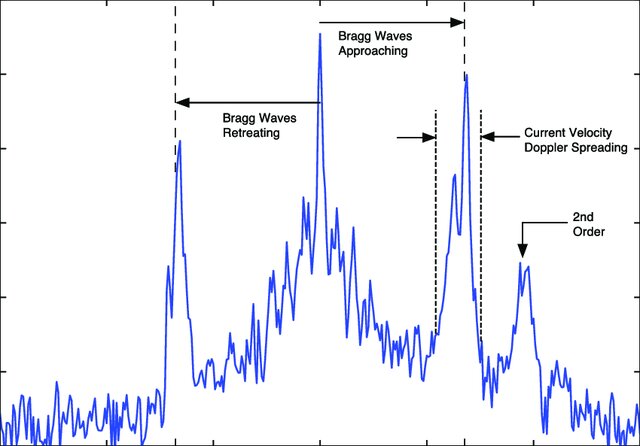}
    \caption{Typical HF-radar ocean-backscatter Doppler spectrum, showing the first-order Bragg peaks and the weaker second-order continuum used in conventional wave spectrum and parameter retrieval.}
    \label{fig:intro_doppler_spectrum}
\end{figure}

Conventional HF-radar wave estimation relies on calculation of the
Doppler spectrum of the received signal.
As illustrated in Fig.~\ref{fig:intro_doppler_spectrum},
the spectrum contains prominent first-order Bragg peaks
and a weaker second-order continuum. The Bragg peaks
arise from resonant scattering by ocean waves with
wavelengths equal to half the radar wavelength,
whereas the second-order return contains information
about a broader range of ocean-wave frequency components.
Established methods estimate SWH through weighted
integration of the second-order return, normalized
by the first-order energy, or through inversion
of the relationship between the radar Doppler
spectrum and the ocean wave spectrum
\cite{Barrick1977,Wyatt1986}.

The weak second-order continuum can be difficult
to distinguish from the noise floor, particularly
when obscured by interference or leakage from
the stronger Bragg peaks. This motivates
time-domain approaches such as the Linear Ordered
Statistic Estimator (LOSE), which maps the largest
rank-ordered magnitudes of Bartlett-beamformed
voltages to SWH through least-squares regression
\cite{shahidi_chandrasekara_linear}.
Despite still being affected by measurement noise,
LOSE avoids explicit isolation of the weak
second-order spectral contribution.

Although LOSE demonstrates the potential of
ordered envelope statistics for time-domain
SWH estimation, its accuracy may be limited
by the fixed spatial processing used to
generate those statistics. Adapting the
spatial weights emphasizes signal
components informative about SWH and suppress
components that contribute to prediction error,
thereby producing ordered-magnitude features
that support more accurate estimation.
The motivation of this research is therefore
to investigate whether incorporating adaptive beamforming
into the structure of the artificial neural network used in the training process can improve
time-domain SWH estimation accuracy beyond
that achieved with fixed Bartlett processing.

This article extends LOSE with three trainable complex
spatial filters with weights iteratively
adjusted using gradient-based optimization
according to the SWH prediction objective.
Ordered magnitudes from their outputs are
concatenated for least-squares SWH estimation.
The adaptive estimator is compared with fixed
Bartlett-based LOSE using common radar--buoy
observations.

\section{Literature Review}
\label{sec:literature_review}

Previously, HF-radar based SWH estimation has been investigated through
Doppler-spectrum analysis and temporal signal statistics.
These approaches differ in the representation of the received HF-radar data used to
relate ocean backscatter to wave height, while spatial
filtering determines the radar signals from which
these representations are constructed.

\subsection{Conventional Frequency-Domain Methods}
\label{subsec:spectral_methods}

Conventional HF-radar ocean surface wave spectrum and parameter retrieval relates the
ocean-backscatter Doppler spectrum to the underlying
ocean-wave spectrum. The first-order Bragg peaks and
second-order continuum illustrated in
Fig.~\ref{fig:intro_doppler_spectrum} provide an example of the spectral
components used in established retrieval methods
\cite{Barrick1977,Wyatt1986}.

Barrick \cite{Barrick1977} developed a method for
extracting wave parameters from weighted integrals
of the second-order Doppler spectrum. Normalization
by the first-order return relates the measured
spectral energy to ocean-wave energy, permitting
bulk wave-parameter estimation without reconstructing
the complete directional spectrum. This approach
requires reliable identification of the first- and second-order scattering
contributions and sufficient second-order signal
strength.

Directional-spectrum inversion recovers a more
detailed description of the wave field. Wyatt
\cite{Wyatt1986} investigated the recovery of
directional ocean-wave spectra from HF-radar
measurements, while Hisaki \cite{Hisaki1996}
developed nonlinear inversion of the relationship
between the ocean and radar spectra. Shahidi and
Gill \cite{ShahidiInverse2021} subsequently introduced
an automatic nonlinear optimization method for
directional-spectrum extraction from monostatic
HF-radar data. Once the sea-surface elevation
frequency spectrum $S_{\eta}(f)$ is obtained,
spectral SWH maybe estimated by: 
\begin{equation}
    H_s = 4\sqrt{m_0},
    \qquad
    m_0 = \int_0^\infty S_{\eta}(f)\,\mathrm{d}f.
    \label{eq:review_spectral_swh}
\end{equation}
where, $m_0$ denotes the zeroth spectral moment,
corresponding to the sea-surface elevation variance.

The applicability of these methods depends on
the scattering approximation and spectral quality.
Wyatt \cite{WyattLimits2000} examined limitations
of HF-radar spectral inversion, and Wyatt et al.\
\cite{WyattQuality2011} investigated the data-quality
requirements for wave measurement. Weak second-order
returns approaching the noise floor, spectral
contamination, and difficulty separating scattering
contributions can restrict retrieval. These limitations
motivate approaches based on temporal signal statistics
that avoid explicit extraction of the second-order
Doppler continuum.

\subsection{Time-Domain Methods}
\label{subsec:temporal_methods}

Time-domain approaches relate SWH to statistical
properties of the received radar signal. Shahidi
and Gill \cite{Shahidi2018} investigated SWH
estimation without constructing Doppler spectra
and subsequently developed two related estimators
\cite{Shahidi2020}. One uses the variance of
short-time Fourier transform coefficients, while
the other operates directly on the standard deviation of
the received time series. These methods establish
calibrated relationships between signal statistics
and SWH under the adopted scattering approximation.

Hashemi et al.\ \cite{Hashemi2023} extended temporal
estimation to synthetic measurements involving
a floating transmitter and a stationary receiver.
Their formulation included a quadratic dependence
of the received field on sea-surface elevation
and used upper-envelope statistics to estimate
SWH. The simulation study reported RMSEs of
approximately 0.10--0.14~m over wave heights
of 0.5--4.5~m, supporting temporal estimation
under the investigated scattering and
platform-motion conditions.

Bragg-wave amplitude modulation provides another
temporal approach. Morales-M\'arquez et al.\
\cite{MoralesMarquez2025} compared this method
with second-order spectral estimation using
coastal HF-radar observations. The spectral
method performed better in their evaluation,
including at longer ranges. Avoiding explicit
Doppler-spectrum construction therefore does
not by itself establish improved retrieval
accuracy; the effectiveness of the temporal
representation must be assessed experimentally.

Chandrasekara and Shahidi
\cite{shahidi_chandrasekara_linear} introduced the Linear
Ordered Statistic Estimator (LOSE), which estimates SWH
from ordered voltage magnitudes. Its physical motivation
builds on the first-order relationship
\begin{equation}
E_{0,n,1}^{+}(\rho,\theta,t) = M(\rho,\theta)\xi(\rho,\theta,t),
\label{eq:spatoi_temp_relationship}
\end{equation}
where $\xi$ denotes sea-surface elevation and
$M(\rho,\theta)$ is a complex-valued spatial
proportionality factor for the ocean patch at coordinates
$(\rho,\theta)$. Taking magnitudes motivates relating
the amplitude statistics of the received voltage to
those of the surface elevation.

Using a Rayleigh envelope model and treating positive
and negative envelope excursions as independent, the
LOSE derivation represents the voltage-magnitude
upper-envelope CDF as
\begin{equation}
    F_{E,\text{upper}}(x; \mu) = \left( 1 - e^{-\frac{K^2 x^2}{2\mu^2}} \right)^2. \label{eq:radar_voltage_cdf}
\end{equation}
Here, $\mu$ is the Rayleigh scale parameter and $K$ is
a positive system scaling constant. Inverting this
CDF gives the rank-based expression used in the
derivation \cite{shahidi_chandrasekara_linear}:
\begin{equation}
    \mathbb{E}\left[ E_{0,n,(\text{upper},k,N)}^+ \right] = \frac{\mu}{K} \sqrt{2 \ln \left( \frac{1}{1 - \sqrt{\frac{k}{N}}} \right)}. \label{eq:final_inverse_cdf}
\end{equation}
The displayed relationship is interpreted as a
rank-based quantile approximation for $k<N$, where
$k$ denotes the ascending rank among $N$ envelope
samples. Its proportionality to $\mu$ motivates
a linear mapping from ordered observations to the
wave-height-related scale. Since $\mu$ is proportional
to the elevation standard deviation $\sigma_\xi$ and
$H_s\approx4\sigma_\xi$, the estimator uses a linear combination of received voltages to estimate the SWH:
\begin{equation}
    H_s = \sum_{j=1}^{M} w_j E_{0,n,(\text{upper},N+1-j,N)}^+. \label{eq:linear_combination_basis}
\end{equation}
Here, $M$ denotes the number of retained highest-ranked
observations and the $w_j$ are the fitted coefficients.
The limit of summation $M$ is distinct from the spatial factor
$M(\rho,\theta)$ in
Equation~\ref{eq:spatoi_temp_relationship}.

In implementation, LOSE substitutes the largest raw
complex-voltage magnitudes for explicitly-extracted
upper-envelope observations. The retained features
are fitted to reference SWH measurements through
least-squares regression, followed by nonnegative
activation and backward temporal averaging.
Evaluation using the Argentia WERA measurements
yielded a reported RMSE of 0.224~m with 181 retained
ranks, 219 training records, and a 25-record averaging
window \cite{shahidi_chandrasekara_linear}. This
ordered-statistical mapping forms the baseline for
the present study.

The Quadratic Second-Order Scatter Compensation
Estimator (QSOCE) extends this framework by introducing
quadratic compensation before rank ordering
\cite{ChandrasekaraQSOCE2026}. Higher-order terms in
the Neumann expansion of the Walsh surface-field
formulation motivate the approximate quadratic model
\cite{Gill1999HFScattering}
\begin{equation}
    E_{0,n}^{+} = A\xi^2 + B\xi + C,
    \label{eq:quadratic_surface}
\end{equation}
where $A$, $B$, and $C$ are system- and
propagation-dependent coefficients. For mean-removed
elevation, $\mathbb{E}[\xi]=0$ and
$\mathbb{E}[\xi^2]\approx H_s^2/16$. Absorbing
$1/16$ into the quadratic coefficient $A$ gives 
\begin{equation}
\mathbb{E}\!\left[E_{0,n}^{+}\right]
\approx
AH_s^2
+
C.
\label{eq:quadratic_theory}
\end{equation}
This statistical relationship motivates an inverse
calibrated quadratic mapping of radar-derived samples.
The compensated samples are rank ordered and processed
through least-squares regression, nonnegative activation,
and causal averaging. Among magnitude, in-phase, and
quadrature representations, the best reported quadrature
configuration achieved an RMSE of 0.187~m and a
correlation coefficient of approximately 0.97
\cite{ChandrasekaraQSOCE2026}.

The Windowed Envelope Statistics Estimator (WESE)
further develops temporal feature extraction by replacing
individual samples with localized statistics
\cite{ChandrasekaraWESE2026}. For a real-valued signal representation $x_i[n]$ from
record $i$, windows of length $L$ and step $D$ yield
the local mean, variance, and standard deviation:
\begin{align}
    \mu_{i,r}
    &=
    \frac{1}{L}\sum_{n=0}^{L-1}x_i[rD+n],
    \label{eq:review_wese_mean}\\
    \upsilon_{i,r}
    &=
    \frac{1}{L}\sum_{n=0}^{L-1}
    \left(x_i[rD+n]-\mu_{i,r}\right)^2,
    \label{eq:review_wese_variance}\\
    \varsigma_{i,r}
    &=
    \sqrt{\upsilon_{i,r}},
    \label{eq:review_wese_std}
\end{align}
where $r$ is the window index. The study compared mean,
standard deviation, and variance features from in-phase,
quadrature, and magnitude sequences. Quadratic
compensation, rank ordering, least-squares regression,
and causal smoothing followed feature extraction.
Its best reported configuration used quadrature variance,
a 16-sample window, 896 retained chronological samples,
and 30-h smoothing, achieving an RMSE of 0.152~m and
a correlation coefficient of 0.978
\cite{ChandrasekaraWESE2026}.

These developments address the temporal representation
and its mapping to SWH. Their reported performance
also incorporates calibration and smoothing choices.
The present study examines the \textit{spatial} combination
that generates the temporal input, using matched
evaluation conditions to assess its effect.

\subsection{Spatial Filtering for Time-Domain SWH Estimation}
\label{subsec:spatial_filtering_review}

Spatial filtering determines the angular contributions
to the radar time series supplied to the estimator.
For an array measurement vector $\mathbf{r}(n)$
and complex weight vector $\mathbf{b}$, the combined
signal is
\begin{equation}
    y(n)=\mathbf{b}^{\mathsf{H}}\mathbf{r}(n).
    \label{eq:review_beamforming}
\end{equation}
Here, $(\cdot)^{\mathsf{H}}$ denotes the Hermitian
(complex conjugate transpose) operator.
The weight amplitudes and phases determine the
directional response \cite{VanVeen1988}. Changing
these weights also changes the magnitude sequence
$|y(n)|$ and its ordered samples, directly affecting
the inputs to LOSE.

Conventional Bartlett beamforming uses the steering
vector $\mathbf{a}(\theta_0)$ for a selected direction
$\theta_0$. With unity-response normalization,
\begin{equation}
    \mathbf{b}_{\mathrm{B}}
    =
    \frac{\mathbf{a}(\theta_0)}
    {\mathbf{a}^{\mathsf{H}}(\theta_0)\mathbf{a}(\theta_0)}.
    \label{eq:review_bartlett}
\end{equation}
For an ideal array with unit-magnitude steering-vector
entries, the weights have equal amplitudes and
direction-dependent phases \cite{VanVeen1988}.
Bartlett beamforming therefore differs from unsteered equal-weight
averaging through its directional phase compensation,
as summarized in Table~\ref{tab:review_beamformers}.

Minimum variance distortionless response (MVDR)
beamforming minimizes output power while preserving
the response in the selected direction $\theta_0$
\cite{VanVeen1988}:
\begin{equation}
    \mathbf{b}_{\mathrm{MVDR}}
    =
    \frac{\mathbf{R}^{-1}\mathbf{a}(\theta_0)}
    {\mathbf{a}^{\mathsf{H}}(\theta_0)
     \mathbf{R}^{-1}\mathbf{a}(\theta_0)},
    \label{eq:review_mvdr}
\end{equation}
where $\mathbf{R}$ is the array covariance matrix.
Its practical performance depends on covariance
estimation and steering accuracy. Vorobyov et al.\
\cite{Vorobyov2003} examined sensitivity to signal
mismatch and developed a robust adaptive formulation.

\begin{table}[!t]
    \centering
    \caption{Comparison of spatial combining methods.
    Equal-weight averaging denotes unsteered combining.}
    \label{tab:review_beamformers}
    \footnotesize
    \renewcommand{\arraystretch}{1.15}
    \setlength{\tabcolsep}{3pt}
    \begin{tabular}{@{}
        >{\raggedright\arraybackslash}p{0.23\columnwidth}
        >{\raggedright\arraybackslash}p{\dimexpr0.77\columnwidth-6pt\relax}
        @{}}
        \hline
        \textbf{Method}
        & \textbf{Characteristics and limitations} \\
        \hline

        Unsteered equal-weight averaging
        & Identical channel weights without directional
          phase compensation or covariance estimation.
          The response is not steered toward an arbitrary
          look direction \cite{VanVeen1988}. \\
        \hline

        Bartlett
        & Steering-vector weights provide directional
          phase compensation, with equal amplitudes for
          an ideal array. No covariance estimate is
          required; weights do not adapt to measured
          interference \cite{VanVeen1988}. \\
        \hline

        MVDR
        & Uses steering and covariance information to
          minimize output power while preserving the
          selected-direction response
          \cite{VanVeen1988}. Sensitive to steering
          mismatch and covariance-estimation errors
          \cite{Vorobyov2003}. \\
        \hline
    \end{tabular}
\end{table}

Bartlett beamforming is selected as the reference because it applies
a fixed, data-independent linear transformation to the
received voltages \cite{VanVeen1988}. For a fixed array
geometry and look direction, its consistent spatial
response supports interpretation of the ordered
magnitudes through the scale relationship underlying
LOSE. MVDR also combines voltages linearly for a given
weight vector, but covariance-dependent adaptation can
change the angular weighting of distributed ocean
backscatter between records, despite preserving the
specified look-direction response.

A fixed spatial response does not establish optimality
for SWH estimation. It provides a controlled baseline
against which spatial weights optimized for the
estimation task can be evaluated. Comparing the proposed
processing with Bartlett-based LOSE under matched
calibration, evaluation, and temporal-smoothing
conditions allows the effect of spatial processing
on SWH accuracy to be assessed.

\section{Methodology}
\label{sec:methodology}

The Linear Ordered Statistic Estimator (LOSE)
maps ordered magnitudes of Bartlett-beamformed
radar voltages to significant wave height (SWH)
through least-squares regression
\cite{shahidi_chandrasekara_linear}.
Its adaptive multi-beam extension incorporates
trainable complex spatial weights into this
framework, allowing the spatial representation
to be adjusted according to the SWH estimation
objective.

Both architectures use magnitude extraction,
rank ordering, least-squares regression, and
causal sliding-window averaging. Bartlett-based
LOSE retains a single spatial output with fixed
beamforming weights, whereas the adaptive model
concatenates ordered features from three spatial
outputs whose complex weights are optimized
during training.

\subsection{Linear Ordered Statistic Estimator}
\label{subsec:method_lose}
\subsubsection{\textbf{Bartlett-Beamformed Input}}

The original LOSE framework operates on complex
radar voltages obtained through Bartlett beamforming
\cite{shahidi_chandrasekara_linear}.
Using the array snapshot $\mathbf{r}(n)$ and
Bartlett weights $\mathbf{b}_{\mathrm{B}}$ defined
in Equation \ref{eq:review_bartlett}, the combined signal is
obtained from
\begin{equation}
    y(n)
    =
    \mathbf{b}_{\mathrm{B}}^{\mathsf{H}}
    \mathbf{r}(n).
    \label{eq:method_lose_beamforming}
\end{equation}
The spatial weights remain fixed for the selected
array geometry and look direction. LOSE therefore
calibrates the mapping from the resulting ordered
voltage magnitudes to reference SWH without
adapting the spatial response.

\subsubsection{\textbf{First-Order Scattering and Ordered Statistics}}

The estimator is motivated by the first-order
scattering relationship
\eqref{eq:spatoi_temp_relationship} and by the
Rayleigh upper-envelope derivation in
Section~\ref{subsec:temporal_methods}.
The rank expression \eqref{eq:final_inverse_cdf}
is proportional to the Rayleigh scale $\mu$.
Under the adopted surface model, $\mu$ is
proportional to the elevation standard deviation
$\sigma_\xi$, with $H_s\approx 4\sigma_\xi$.
This leads to the linear combination of the
largest ordered amplitudes in
\eqref{eq:linear_combination_basis}, where $M$
is the number of retained highest-ranked
observations and $w_j$ are calibrated
coefficients. Once again, the integer $M$ is distinct from
the spatial factor $M(\rho,\theta)$.

In this implementation, the largest raw
complex-voltage magnitudes are used in place of
explicitly-extracted upper-envelope observations
\cite{shahidi_chandrasekara_linear}.
The coefficients $w_j$ are obtained by the
least-squares fit in the next subsection.

\subsubsection{\textbf{Least-Squares Model}}

For each acquisition interval, the voltage
magnitudes are rank ordered and the largest
$M$ values are retained. These features form
a row of the radar feature matrix
$\mathbf{E}\in\mathbb{R}^{N_{\mathrm{tr}}\times M}$,
where $N_{\mathrm{tr}}$ denotes the number of
calibration observations. The corresponding
reference SWH measurements form the vector
$\mathbf{h}_{\mathrm{tr}}
\in\mathbb{R}^{N_{\mathrm{tr}}}$.
The regression vector $\mathbf{w}_{\mathrm{opt}}$ is defined by
\begin{equation}
    \mathbf{w}_{\mathrm{opt}}
    =
    \arg\min_{\mathbf{w}}
    \left\|
        \mathbf{E}\mathbf{w}
        -
        \mathbf{h}_{\mathrm{tr}}
    \right\|_2^2.
    \label{eq:method_least_squares_objective}
\end{equation}
The minimum-norm least-squares solution is
\begin{equation}
    \mathbf{w}_{\mathrm{opt}}
    =
    \mathbf{E}^{\dagger}
    \mathbf{h}_{\mathrm{tr}},
    \label{eq:method_least_squares_pseudoinverse}
\end{equation}
where $(\cdot)^{\dagger}$ denotes the
Moore--Penrose pseudoinverse \cite{Penrose1955}. The estimates
for the feature matrix are
\begin{equation}
    \hat{\mathbf{H}}_s
    =
    \mathbf{E}\mathbf{w}_{\mathrm{opt}}.
    \label{eq:method_linear_swh_estimate}
\end{equation}
The fitted coefficients are subsequently applied
to features from new acquisition intervals.

\begin{figure*}[!t]
    \centering
    \includegraphics[width=2.0\columnwidth]{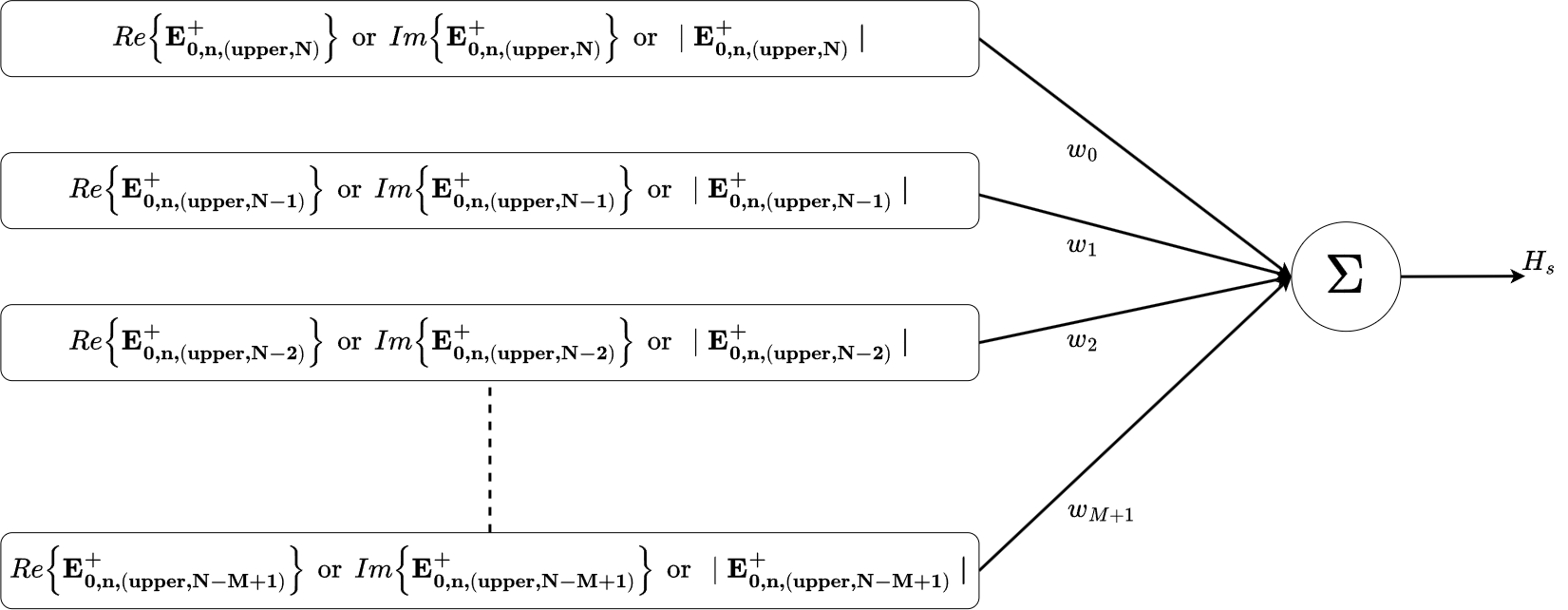} 
    \caption{Bartlett-based LOSE architecture. The largest $M$ magnitudes, real, or imaginary parts of the received voltages are retained after fixed spatial filtering and mapped to SWH through least-squares regression. The resulting estimates are smoothed using causal sliding-window averaging.}
    \label{fig:method_lose_architecture}
\end{figure*}

The LOSE architecture is illustrated in
Fig.~\ref{fig:method_lose_architecture}.
The spatial weights are fixed, whereas the
ordered-feature regression coefficients are
determined from reference SWH measurements.

\subsection{Adaptive Multi-Beam Complex Spatial Filtering}
\label{subsec:method_adaptive}

The adaptive spatial extension retains the
ordered-statistical regression structure of
LOSE while incorporating the complex spatial
weights into the SWH estimation model.
Multiple spatial filters produce separate
voltage sequences, from which ordered magnitude
features are extracted and concatenated.

The spatial weights influence the features
supplied to regression and are adjusted according
to the resulting SWH prediction error.
The architecture therefore extends calibration
to the spatial combination of the receiver
channels.

\subsubsection{Complex Multi-Beam Formation}

Let the array snapshot at sample $n$ during
acquisition interval $i$ be
\begin{equation}
    \mathbf{y}_i[n]
    =
    \begin{bmatrix}
        y_{i,1}[n] &
        y_{i,2}[n] &
        \cdots &
        y_{i,M_a}[n]
    \end{bmatrix}^{T},
    \label{eq:method_adaptive_array_snapshot}
\end{equation}
where $M_a$ is the number of receiver channels.
For a uniform linear array, the steering vector
for nominal direction $\theta_b$ is
\begin{equation}
    \mathbf{a}(\theta_b)
    =
    \begin{bmatrix}
        1 &
        e^{j2\pi d\sin(\theta_b)} &
        \cdots &
        e^{j2\pi d(M_a-1)\sin(\theta_b)}
    \end{bmatrix}^{T}.
    \label{eq:method_adaptive_steering_vector}
\end{equation}
Here, $d$ is the inter-element spacing expressed
in wavelengths, $\theta_b$ is measured relative
to array broadside, and $j=\sqrt{-1}$. The output of the $b^{th}$ spatial filter is
\begin{equation}
    z_{i,b}[n]
    =
    \mathbf{w}_b^{H}\mathbf{y}_i[n],
    \qquad b=1,2,\ldots,B,
    \label{eq:method_adaptive_beam_output}
\end{equation}
where $\mathbf{w}_b\in\mathbb{C}^{M_a}$ is its
complex spatial-weight vector and $B$ is the
number of filters.

The spatial vectors are initialized using
steering vectors around the buoy look direction.
For a symmetric three-beam arrangement,
\begin{equation}
    \theta_b
    \in
    \left\{
        \theta_0-\Delta\theta,\,
        \theta_0,\,
        \theta_0+\Delta\theta
    \right\},
    \label{eq:method_adaptive_initial_angles}
\end{equation}
where $\theta_0$ is the buoy look angle and
$\Delta\theta$ is the angular offset.

Unlike the fixed Bartlett weights in LOSE,
the complex spatial vectors are learnable.
Their amplitudes and phases can therefore
change from the initial steering configuration.
The resulting spatial responses need not remain
centered on their initial directions.

\subsubsection{Multi-Beam Ordered Features}
Each spatial output is converted to a magnitude
sequence:
\begin{equation}
    e_{i,b}[n]
    =
    |z_{i,b}[n]|.
    \label{eq:method_adaptive_envelope}
\end{equation}
The samples from each beam are independently
rank ordered:
\begin{equation}
    e_{i,b,(N)}
    \ge
    e_{i,b,(N-1)}
    \ge\cdots\ge
    e_{i,b,(1)}.
    \label{eq:method_adaptive_ordering}
\end{equation}
Retaining the largest $M$ values gives the vector
\begin{equation}
    \mathbf{u}_{i,b}
    =
    \begin{bmatrix}
        e_{i,b,(N)} &
        e_{i,b,(N-1)} &
        \cdots &
        e_{i,b,(N-M+1)}
    \end{bmatrix}^{T}.
    \label{eq:method_adaptive_top_features}
\end{equation}
The feature vectors are concatenated:

\begin{equation}
    \mathbf{u}_i
    =
    \begin{bmatrix}
        \mathbf{u}_{i,1} &
        \mathbf{u}_{i,2} &
        \cdots &
        \mathbf{u}_{i,B}
    \end{bmatrix}
    \in\mathbb{R}^{BM}.
    \label{eq:method_adaptive_concatenation}
\end{equation}

This construction retains the LOSE principle
of using upper-ranked voltage magnitudes,
but applies it separately to multiple spatial
outputs. Equal ranks across beams need not
correspond to the same temporal sample,
because each magnitude sequence is ordered
independently.

\subsubsection{Least-Squares SWH Mapping}

The concatenated features produce the raw
SWH estimate
\begin{equation}
    \hat{H}_{s,i}
    =
    \mathbf{u}_i^{T}\boldsymbol{\beta},
    \label{eq:method_adaptive_prediction}
\end{equation}
where $\boldsymbol{\beta}\in\mathbb{R}^{BM}$
contains the regression coefficients.

For a given set of spatial weights, the
feature vectors form the matrix
\begin{equation}
    \mathbf{U}
    =
    \begin{bmatrix}
        \mathbf{u}_1^{T}\\
        \mathbf{u}_2^{T}\\
        \vdots\\
        \mathbf{u}_{N_{\mathrm{tr}}}^{T}
    \end{bmatrix}
    \in\mathbb{R}^{N_{\mathrm{tr}}\times BM}.
    \label{eq:method_adaptive_feature_matrix}
\end{equation}
With corresponding buoy observations collected
in $\mathbf{h}\in\mathbb{R}^{N_{\mathrm{tr}}}$,
the regression solution is
\begin{equation}
    \boldsymbol{\beta}
    =
    \mathbf{U}^{\dagger}\mathbf{h}.
    \label{eq:method_adaptive_ls_solution}
\end{equation}
This minimizes
\begin{equation}
    \left\|
        \mathbf{U}\boldsymbol{\beta}
        -
        \mathbf{h}
    \right\|_2^2
    \label{eq:method_adaptive_ls_error}
\end{equation}
for the specified spatial weights.

The adaptive component arises because changes
to $\mathbf{w}_b$ modify the beamformed signals,
their ordered magnitude features, and therefore
the matrix $\mathbf{U}$. The spatial weights
are adjusted according to SWH prediction error,
while the regression coefficients are obtained
through least-squares mapping for the associated
features. The numerical training procedure
is described in the experimental evaluation.

\begin{figure*}[!t]
    \centering
    \includegraphics[width=0.8\textwidth]{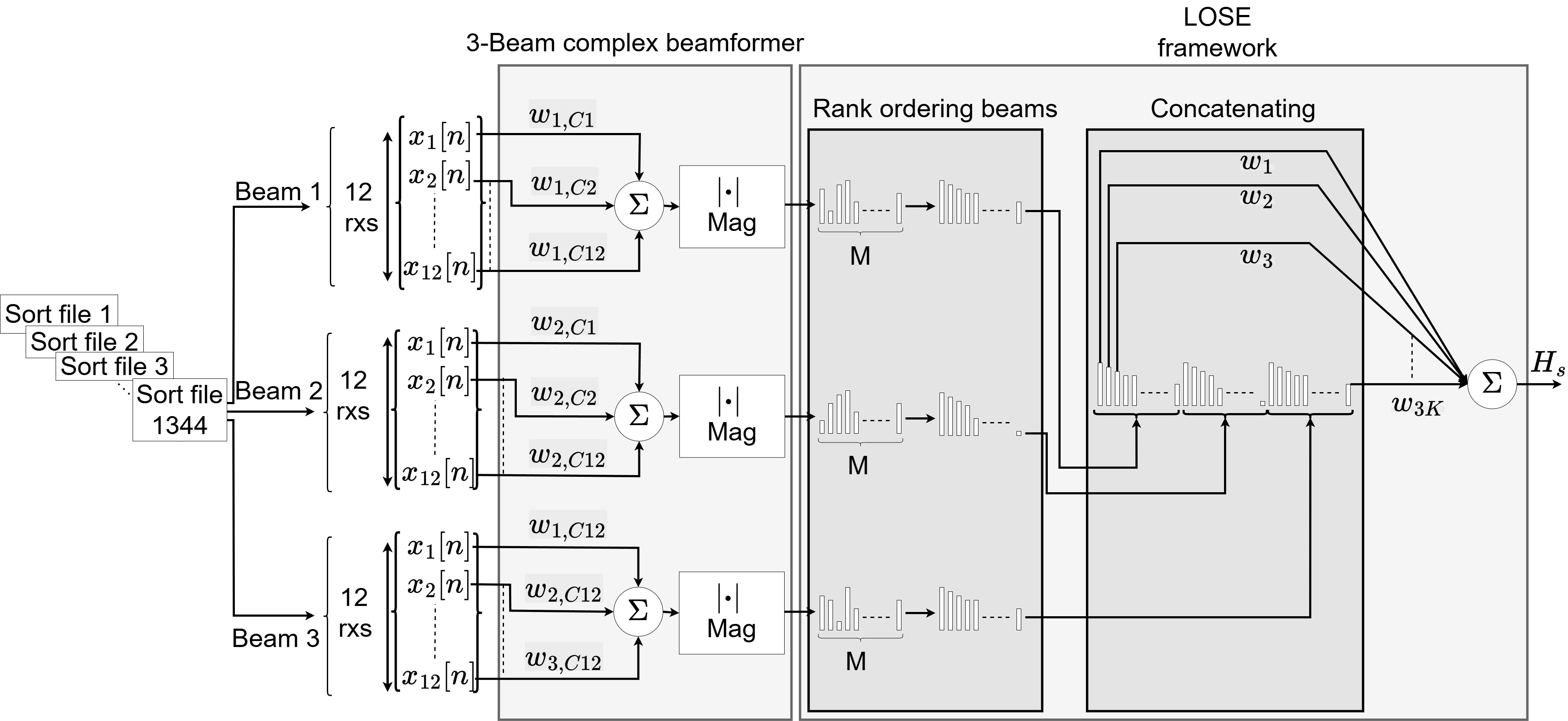}
    \caption{Adaptive multi-beam extension of LOSE. Each learnable spatial filter produces a voltage sequence whose largest $M$ magnitudes are retained. The beamwise features are concatenated and mapped to SWH using least-squares regression. The nonnegative estimates are smoothed using the same causal sliding-window mechanism as LOSE.}
    \label{fig:method_adaptive_architecture}
\end{figure*}

The architecture is illustrated in
Fig.~\ref{fig:method_adaptive_architecture}.
Both this model and the Bartlett-based LOSE
baseline apply nonnegative activation and
causal sliding-window averaging, as described
in Section~\ref{subsec:method_common_smoothing}.

\subsection{Relationship Between the Architectures}
\label{subsec:method_model_comparison}

The two models share magnitude extraction,
ordered-feature selection, and linear
least-squares mapping. Their primary distinction
is the treatment of spatial processing.
In the original LOSE model, Bartlett weights
are fixed before regression calibration.
In the adaptive model, the complex spatial
weights are learned according to the SWH
prediction objective, allowing the input
features to change during training.

The multi-beam architecture additionally
increases the feature dimension from $M$
to $BM$. Thus, the extension incorporates
both spatial-weight learning and a representation
formed from multiple spatial outputs.
The original structure is recovered by using
one fixed Bartlett beam and its associated
ordered features.

For any given spatial-weight vector, beamforming
remains a linear combination of complex radar
voltages. The subsequent magnitude extraction
and ordering are nonlinear operations, while
the final regression remains linear in the
constructed features. The architecture therefore
fits within a neural-network framework:
complex beamforming forms a trainable linear
layer, magnitude extraction and ordered-feature
selection provide nonlinear transformations,
and least-squares regression forms the linear
output layer. In this implementation, the
spatial weights are updated using gradient-based
optimization, while the output coefficients
are obtained through least-squares fitting.

\subsection{Common Nonnegative Activation and Temporal Smoothing}
\label{subsec:method_common_smoothing}

The raw estimates from both architectures are
constrained to be nonnegative before temporal
smoothing, which is equivalent to the application of a ReLU activation function. A causal sliding-window average is then applied using the same mechanism
for LOSE and the adaptive spatial model:
\begin{equation}
    \bar{H}_s[k]
    =
    \frac{1}{P}
    \sum_{i=0}^{P-1}
    \hat{\mathbf{H}}_s[k-i].
    \label{eq:method_moving_average_swh}
\end{equation}
Here, $\hat{\mathbf{H}}_s[k]$ denotes the
nonnegative estimate at acquisition index $k$,
$P$ is the number of estimates in the averaging
window, and $\bar{H}_s[k]$ is the smoothed
output. The expression applies when the
required prediction history is available.

The window moves forward with each acquisition
interval and includes only the current estimate
and preceding estimates. This suppresses rapid
prediction fluctuations while introducing
a temporal averaging effect that can model the delayed
response to changing sea conditions.

The adaptive model uses this same sliding-window
averaging structure. The averaging horizon,
initialization, and treatment of missing
intervals are specified in the experimental
evaluation section.

\section{Experimental Evaluation}
\label{sec:experimental_evaluation}

The effect of incorporating spatial weights into
SWH estimation was evaluated by comparing the
adaptive multi-beam model with Bartlett-based
LOSE using field measurements from Argentia,
Canada. Both models were
calibrated and evaluated against synchronized
buoy observations under a common chronological
data partition and sliding-window smoothing
procedure.

\subsection{Experimental Site and Radar--Buoy Dataset}
\label{subsec:experimental_site}

The measurements were acquired using a shore-based
monostatic WERA HF radar system manufactured by Helzel
and deployed at Argentia, Newfoundland and Labrador,
Canada, at
$(47.3036^{\circ}\mathrm{N},
53.9961^{\circ}\mathrm{W})$.
The radar operated at $13.385~\mathrm{MHz}$,
corresponding to a wavelength of approximately
$22.41~\mathrm{m}$. Its receiving array comprised of
$M_a = 12$ vertical monopole antennas with
an inter-element spacing of $0.45\lambda_0$
(approximately $10.08~\mathrm{m}$).
Each acquisition collected 2,048 complex slow-time
IQ samples for each of the 12 receiving channels
and 40 range gates, with an approximate sampling
interval of $0.39~\mathrm{s}$. Acquisition of each
2,048-sample block lasted approximately
$13.3~\mathrm{min}$, and a new acquisition began
every 15 minutes. The range resolution was
$3.0~\mathrm{km}$.

\begin{figure}[!t]
    \centering
    \includegraphics[width=0.65\columnwidth]{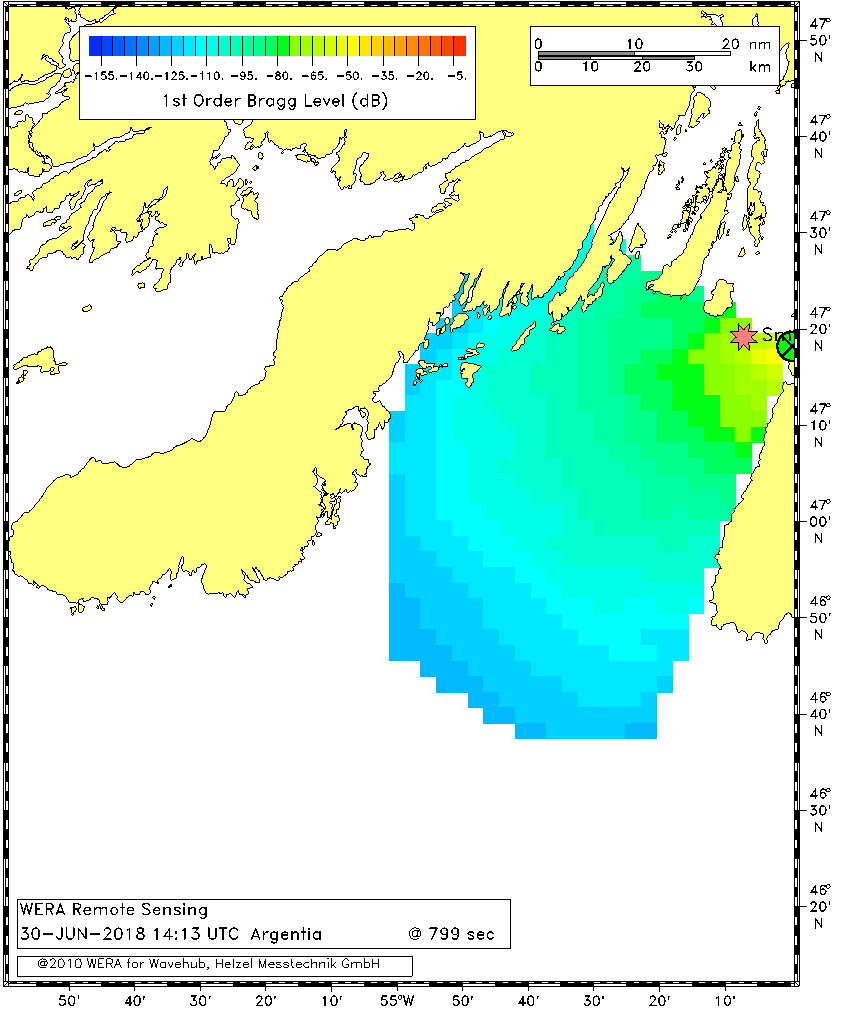} 
    \caption{Locations of the Argentia HF radar and the Red Island Shoal reference wave buoy in Newfoundland and Labrador, Canada.}
    \label{fig:exp_radar_buoy_map}
\end{figure}

Reference SWH measurements were obtained every
15 minutes from the Red Island Shoal directional
wave buoy, initially located at
$(47.3189^{\circ}\mathrm{N},
54.1228^{\circ}\mathrm{W})$. Figure~\ref{fig:exp_radar_buoy_map}
shows the radar and initial buoy locations.

\begin{figure}[!t]
    \centering
    \includegraphics[width=\columnwidth]{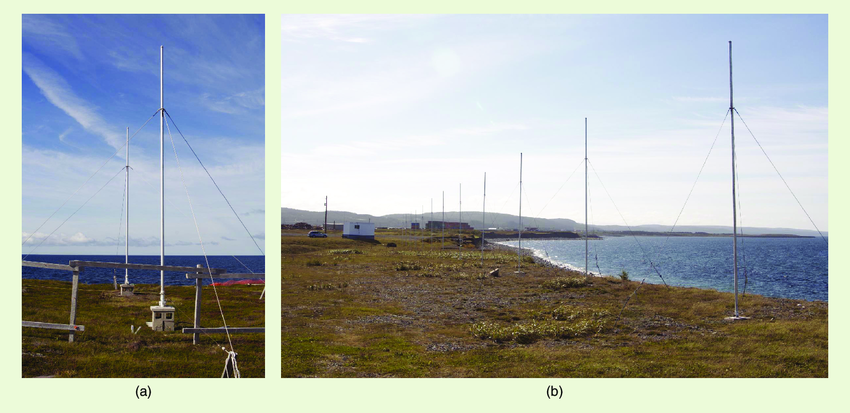}
    \caption{Antenna arrays at the Argentia HF radar
    site: (a) two-element transmitting array and
    (b) twelve-element receiving uniform linear array.}
    \label{fig:exp_array_photo}
\end{figure}

The synchronized receiver channels preserved
the relative amplitudes and phases required for
digital spatial-processing. The radar provided measurements over 40 range
gates with a range resolution of
$3.0~\mathrm{km}$. The received signals were
recorded as complex in-phase and quadrature
(IQ) baseband voltages. A photograph of the
receiving array is shown in
Fig.~\ref{fig:exp_array_photo}, and the principal
system specifications are summarized in
Table~\ref{tab:exp_system_parameters}.

\begin{table}[!t]
    \centering
    \caption{Radar specifications and observation archive.}
    \label{tab:exp_system_parameters}
    \footnotesize
    \renewcommand{\arraystretch}{1.15}
    \setlength{\tabcolsep}{3pt}
    \begin{tabular}{@{}
        >{\raggedright\arraybackslash}p{0.43\columnwidth}
        >{\raggedright\arraybackslash}p{\dimexpr0.57\columnwidth-6pt\relax}
        @{}}
        \hline
        \textbf{Parameter} & \textbf{Specification} \\
        \hline
        Radar system & Monostatic WERA HF radar \\
        Operating frequency & $13.385~\mathrm{MHz}$ \\
        Radar wavelength & Approximately $22.41~\mathrm{m}$ \\
        Receiving array & 12-element monopole ULA \\
        Inter-element spacing
        & $0.45\lambda_0$, approximately $10.08~\mathrm{m}$ \\
        Number of range gates & 40 \\
        Range resolution & $3.0~\mathrm{km}$ \\
        Observation span & 336 h (14 days) \\
        Archive format & Complex IQ measurements in
        \texttt{.SORT} files \\
        Nominal file interval & 15 min \\
        Raw file dimensions
        & $40\times12\times2048$ complex samples \\
        Reference instrument
        & Red Island Shoal directional wave buoy \\
        \hline
    \end{tabular}
\end{table}

\subsection{Data Preparation and Common Evaluation Protocol}
\label{subsec:exp_data_preparation}
\subsubsection{Spatial Alignment with the Buoy}

The nominal buoy coordinates were converted into
range and azimuth relative to the radar site.
The azimuth was then expressed relative to the
receiving-array boresight using the true-north
orientation stored in the radar metadata.
The range mapping included the $0.2~\mathrm{km}$
offset and range-cell centering correction used
in the preprocessing implementation.

\begin{figure}[!t]
    \centering
    \includegraphics[width=1.0\columnwidth]{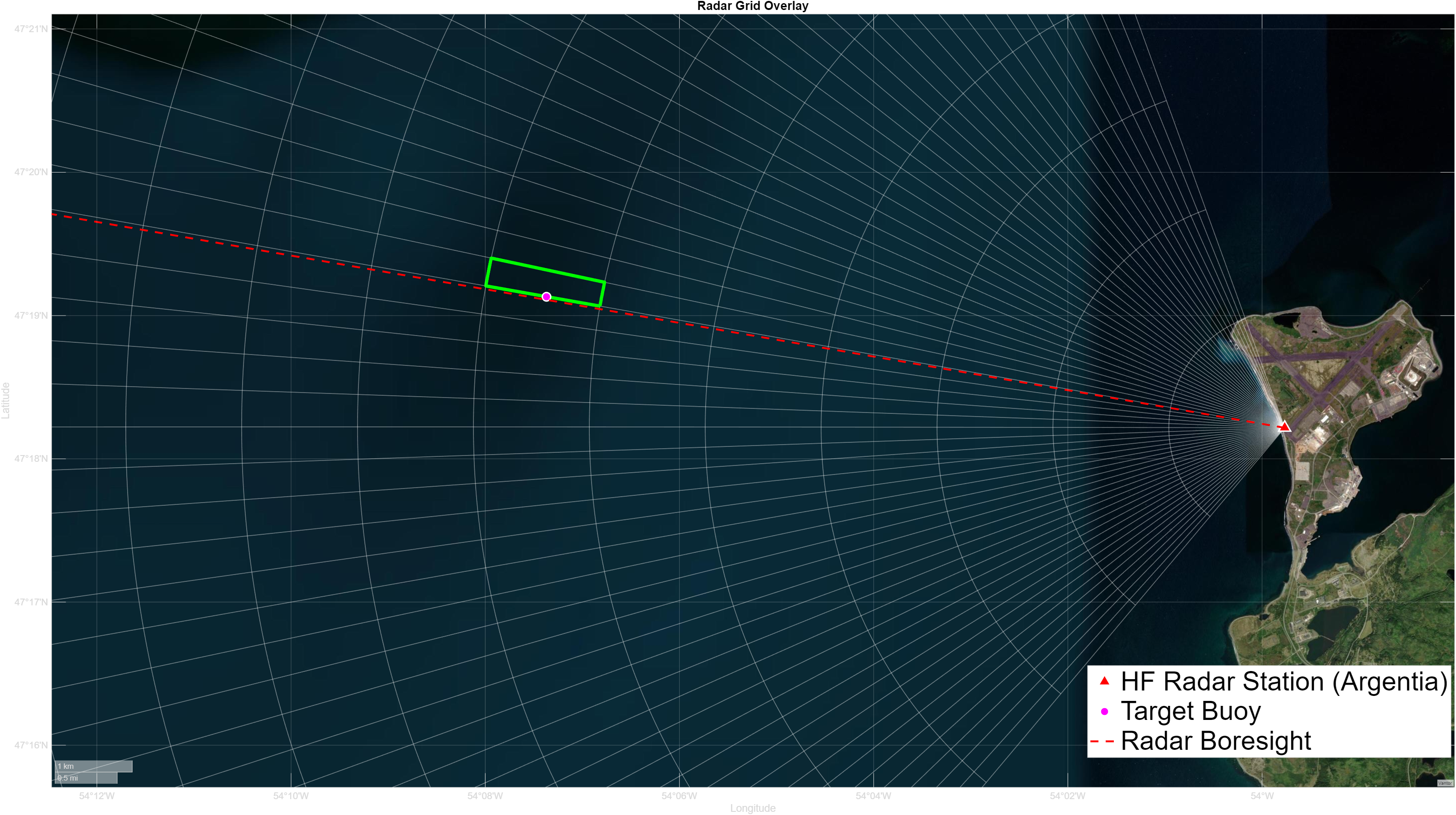} 
    \caption{Radar range--azimuth grid relative to the reference buoy. The overlay identifies the spatial cells used to associate radar observations with the buoy location.}
    \label{fig:exp_radar_grid}
\end{figure}

This mapping identified the range--azimuth
locations associated with the buoy. The spatial
relationship between the radar grid and buoy
is illustrated in Fig.~\ref{fig:exp_radar_grid}.
The selected radar measurements represent a
finite spatial footprint around the buoy rather
than a point measurement at its exact location.

\subsubsection{Radar--Buoy Synchronization and Data Partition}

Each 15-minute radar acquisition contained 2,048
complex samples per receiving channel and was
paired by timestamp with the corresponding buoy
SWH reading. Intervals without valid reference
measurements were excluded. Both processing paths
used the same paired observation intervals.

The observations were ordered chronologically,
with the first 66\% (roughly two-thirds) used for training and the
remaining 34\% (roughly one-third) for evaluation, without random
shuffling. Regression coefficients and adaptive
spatial weights were fitted using only the
training observations. Candidate feature counts
and sliding-window lengths were compared using
post-training RMSE, and the lowest-error
configuration was selected. The evaluation
interval therefore also served as optimal parameter
selection.

\subsection{Bartlett-Based LOSE Implementation and Training}
\label{subsec:exp_lose_implementation}
\subsubsection{\textbf{Fixed Bartlett Beamforming}}

The LOSE baseline used the same multichannel
observations and selected range gate as the
adaptive model. For observation $i$ and temporal
sample $n$, let
$\mathbf{y}_{i,r_c}[n]\in\mathbb{C}^{M_a}$
denote the receiver-channel snapshot at the
selected range gate $r_c$. Fixed Bartlett
beamforming toward the buoy look direction
$\theta_0$ produced
\begin{equation}
    z_{i,\mathrm{b}}[n]
    =
    \mathbf{w}_{\mathrm{B}}^{H}(\theta_0)
    \mathbf{y}_{i,r_c}[n],
    \label{eq:exp_bartlett_beamforming}
\end{equation}
where $\mathbf{w}_{\mathrm{B}}(\theta_0)$
is the unity-response Bartlett weight vector
defined in
Equation~\eqref{eq:review_bartlett}.
The weights remained fixed throughout training
and evaluation. Beamforming was applied directly
to the extracted complex channel measurements,
without neighboring-cell interpolation, producing
one voltage sequence per observation.

\subsubsection{\textbf{Ordered Features and Regression Fitting}}
The beamformed voltage magnitudes were sorted
in descending order, and the largest $M$ values
were retained for each observation. A separate
least-squares model was fitted for each candidate
$M$ using the training observations and their
corresponding buoy SWH measurements. Regression
coefficients were computed using
the python function \texttt{numpy.linalg.lstsq} with
\texttt{rcond=None}, without an intercept or
explicit inversion of the normal-equation matrix.

\subsubsection{\textbf{Sliding-Window Smoothing and Selection}}
The regression predictions were processed using
the causal sliding-window average defined in
the methodology. Smoothing was applied to the
evaluation sequence, with unavailable preceding
values initialized by repeating its first
prediction. No training predictions were carried
into the averaging window.

Candidate feature counts $M$ and averaging
horizons $P$ were compared using post-training
RMSE, and the configuration with the lowest
smoothed RMSE was selected. The evaluation
interval therefore also served as optimal configuration
selection.

\subsection{Adaptive Multi-Beam Implementation and Training}
\label{subsec:exp_adaptive_implementation}
\subsubsection{\textbf{Multichannel Input Preparation}}

The adaptive processing path retained the
complex measurements from the receiver
channels before conventional beamforming.
For acquisition interval $i$, the raw
file contents were represented by
\begin{equation}
    \mathbf{Y}_{i,\mathrm{file}}
    \in
    \mathbb{C}^{N_R\times M_a\times N_S},
    \label{eq:exp_raw_file_tensor}
\end{equation}
where $N_R=40$ is the number of range gates,
$M_a=12$ is the number of receiver channels,
and $N_S=2048$ is the number of temporal
samples per channel in the raw file.

The reference implementation extracted the
range gate associated with the buoy:
\begin{equation}
    \mathbf{Y}_i
    =
    \mathbf{Y}_{i,\mathrm{file}}[r_c,:,:]
    \in\mathbb{C}^{M_a\times N_S},
    \label{eq:exp_target_range_extraction}
\end{equation}
where $r_c$ is the selected range-gate index.
This preserved the channel amplitude and
phase relationships required for adaptive
spatial filtering. The extracted records
were then assembled into the final
observation intervals. Each 15-minute observation
therefore formed a $12\times2048$ complex-valued
array containing 2,048 synchronized samples from
each of the 12 receiving channels.

\subsubsection{\textbf{Spatial-Layer Initialization}}

The neural-network model is comprised of a complex spatial layer and
a linear regression layer, separated by magnitude
extraction and ordered-feature construction.
The spatial layer produces three beam outputs,
with complex weights initialized from steering
vectors directed around the buoy look direction:
\begin{equation}
    \theta_b
    \in
    \left\{
        \theta_0-\Delta\theta,\,
        \theta_0,\,
        \theta_0+\Delta\theta
    \right\},
    \label{eq:exp_initial_beam_angles}
\end{equation}
where $\theta_0$ is the buoy bearing relative
to array broadside and
$\Delta\theta=2.5^{\circ}$.
The angular separation was fixed during
initialization. Subsequent optimization adjusted
the complex channel weights directly, allowing
the spatial responses to depart from their
initial steering configurations.

\subsubsection{\textbf{Spatial-Layer Optimization and Regression Fitting}}

For each observation, the magnitudes of the
three spatial-layer outputs were calculated
and rank ordered independently. The largest
$M$ values from each output were retained and
concatenated into a $3M$-dimensional feature
vector. This vector formed the input to the
linear regression layer.

A separate model was initialized for each
candidate number of retained ranks per beam,
with the full search spanning $M=1,\ldots,N$,
where $N$ is the number of temporal samples
per observation. At each epoch, the current spatial weights
were used to construct the training feature
matrix $\mathbf{U}$. The regression-layer
coefficients were then recalculated using
the same least-squares formulation as LOSE:
\begin{equation}
    \boldsymbol{\beta}
    =
    \mathbf{U}^{\dagger}\mathbf{h},
    \label{eq:exp_regression_layer_fit}
\end{equation}
where $\mathbf{h}$ contains the training
buoy SWH observations and
$\boldsymbol{\beta}\in\mathbb{R}^{3M}$.
The resulting SWH prediction error provided
the objective for updating the complex
spatial weights.

The Adam scheme therefore optimized the spatial layer,
while the regression layer was fitted through
a least-squares solution. After the final spatial
update, the regression coefficients were
recomputed using the resulting training
features. Both sets of parameters were then
held fixed during post-training evaluation.

\subsubsection{\textbf{Sliding-Window Smoothing and Selection}}

The adaptive predictions were constrained to be
nonnegative and smoothed using the same causal
sliding-window averaging procedure as LOSE, with
consistent initialization, handling of missing
intervals, and evaluation timestamps. During the
feature-count sweep, RMSE was recorded before
and after smoothing to distinguish the estimator's
performance from the effect of temporal averaging.
The configuration yielding the lowest post-training
RMSE under the specified smoothing conditions
was selected.

\subsection{Performance Metrics}
\label{subsec:exp_metrics}

Performance was assessed using paired
radar estimates and buoy observations
over the common evaluation interval.
Let $N_e$ denote the number of evaluated
pairs, $H_{s,i}$ the buoy SWH, and
$\tilde{H}_{s,i}$ the corresponding model
prediction. The prediction stage used
for each metric is identified explicitly
as raw, nonnegative unsmoothed, or
sliding-window smoothed.

The root-mean-square error (RMSE) is defined as
\begin{equation}
    \mathrm{RMSE}
    =
    \sqrt{
        \frac{1}{N_e}
        \sum_{i=1}^{N_e}
        \left(
            \tilde{H}_{s,i}-H_{s,i}
        \right)^2
    }.
    \label{eq:exp_rmse}
\end{equation}
RMSE is reported in meters and provides
the primary measure of estimation error.

Temporal agreement is characterized
using the Pearson correlation coefficient:
\begin{equation}
    r
    =
    \frac{
        \sum_{i=1}^{N_e}
        (\tilde{H}_{s,i}-\mu_{\mathrm{p}})
        (H_{s,i}-\mu_{\mathrm{b}})
    }{
        \sqrt{
            \sum_{i=1}^{N_e}
            (\tilde{H}_{s,i}-\mu_{\mathrm{p}})^2
        }
        \sqrt{
            \sum_{i=1}^{N_e}
            (H_{s,i}-\mu_{\mathrm{b}})^2
        }
    },
    \label{eq:exp_correlation}
\end{equation}
where $\mu_{\mathrm{p}}$ and $\mu_{\mathrm{b}}$
are the evaluation-period means of the
predictions and buoy observations,
respectively.

\section{Results and Discussion}
\label{sec:results_discussion}

The adaptive multi-beam estimator reduced the
smoothed SWH RMSE from $0.224~\mathrm{m}$
for Bartlett--LOSE ($M=181$) to
$0.192~\mathrm{m}$ ($M=349$ per beam),
a $14.3\%$ reduction, while preserving the
reported Pearson correlation on the common
evaluation record. Both estimators used
least-squares mapping of ordered voltage
magnitudes and causal sliding-window averaging.
The comparison reflects the selected configurations,
including differences in spatial processing,
feature counts, and averaging windows.

\subsection{Estimation Performance}
\label{subsec:results_temporal}

Table~\ref{tab:results_accuracy_comparison}
summarizes the estimation performance before
and after smoothing. The adaptive configuration
reduced the raw RMSE from $0.912~\mathrm{m}$
to $0.394~\mathrm{m}$, a reduction of
approximately $56.8\%$. Following causal
sliding-window averaging, the RMSE decreased
to $0.224~\mathrm{m}$ for Bartlett--LOSE
and $0.192~\mathrm{m}$ for the adaptive
estimator, a relative reduction of
approximately $14.3\%$. Both smoothed
estimators achieved a Pearson correlation
coefficient of $0.949$ at the reported
precision. The adaptive estimator obtained
its lower smoothed RMSE using a shorter
averaging window of $12$ samples, compared
with $25$ samples for Bartlett--LOSE.

\begin{table}[!t]
    \centering
    \caption{Estimation performance before
    and after causal sliding-window averaging.}
    \label{tab:results_accuracy_comparison}
    \footnotesize
    \renewcommand{\arraystretch}{1.15}
    \setlength{\tabcolsep}{3pt}
    \begin{tabular}{@{}
        >{\raggedright\arraybackslash}p{0.46\columnwidth}
        >{\centering\arraybackslash}p{\dimexpr0.27\columnwidth-6pt\relax}
        >{\centering\arraybackslash}p{\dimexpr0.27\columnwidth-6pt\relax}
        @{}}
        \hline
        \textbf{Quantity}
        & \textbf{Bartlett--LOSE}
        & \textbf{Adaptive} \\
        \hline
        Raw RMSE (m) & 0.912 & 0.394 \\
        Smoothed RMSE (m) & 0.224 & 0.192 \\
        Smoothed Pearson correlation & 0.949 & 0.949 \\
        Smoothing window (samples) & 25 & 12 \\
        \hline
    \end{tabular}
\end{table}

\begin{figure}[!t]
    \centering
    \includegraphics[width=\columnwidth]
    {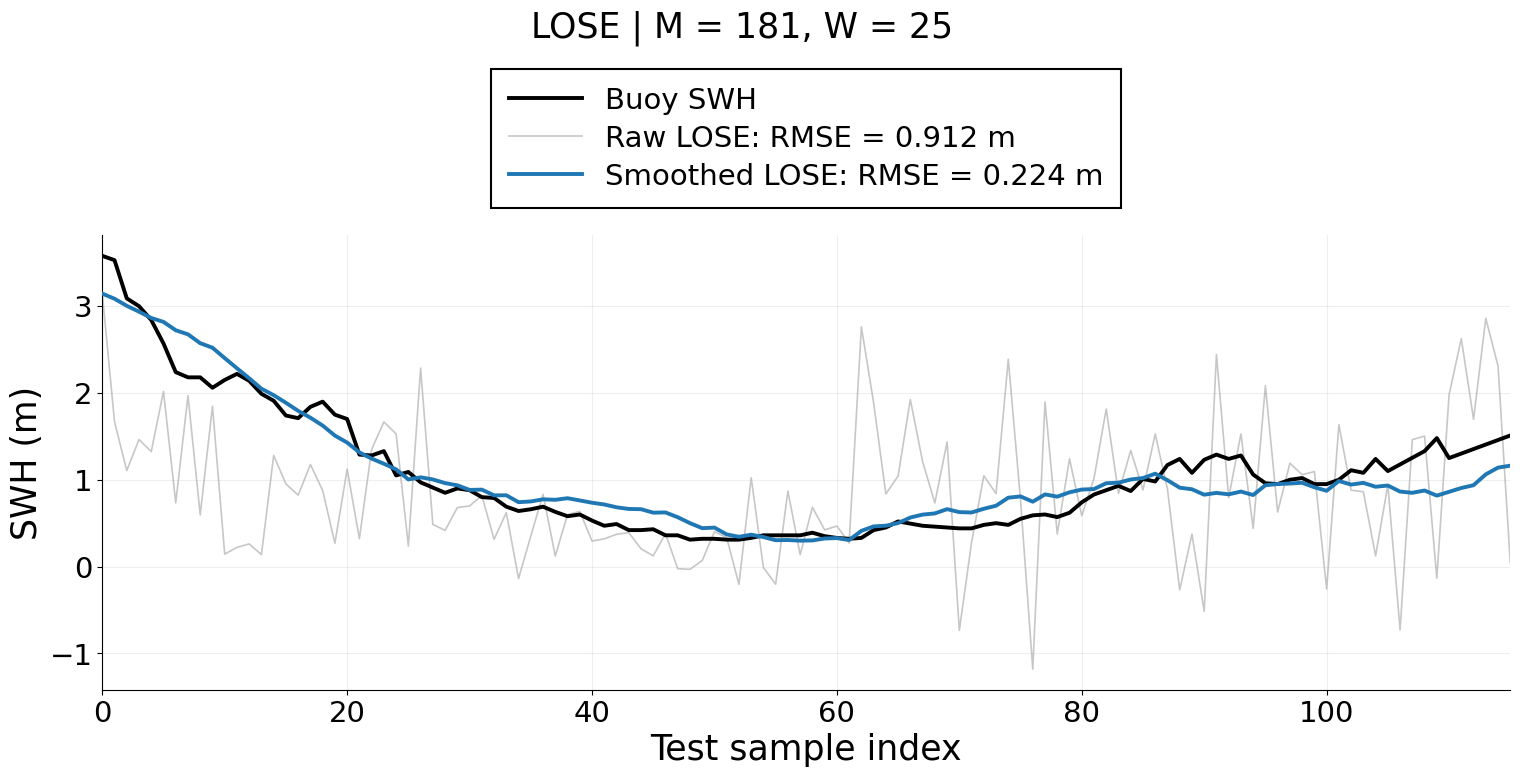}

    \par\vspace{0.5em}

    \includegraphics[width=\columnwidth]
    {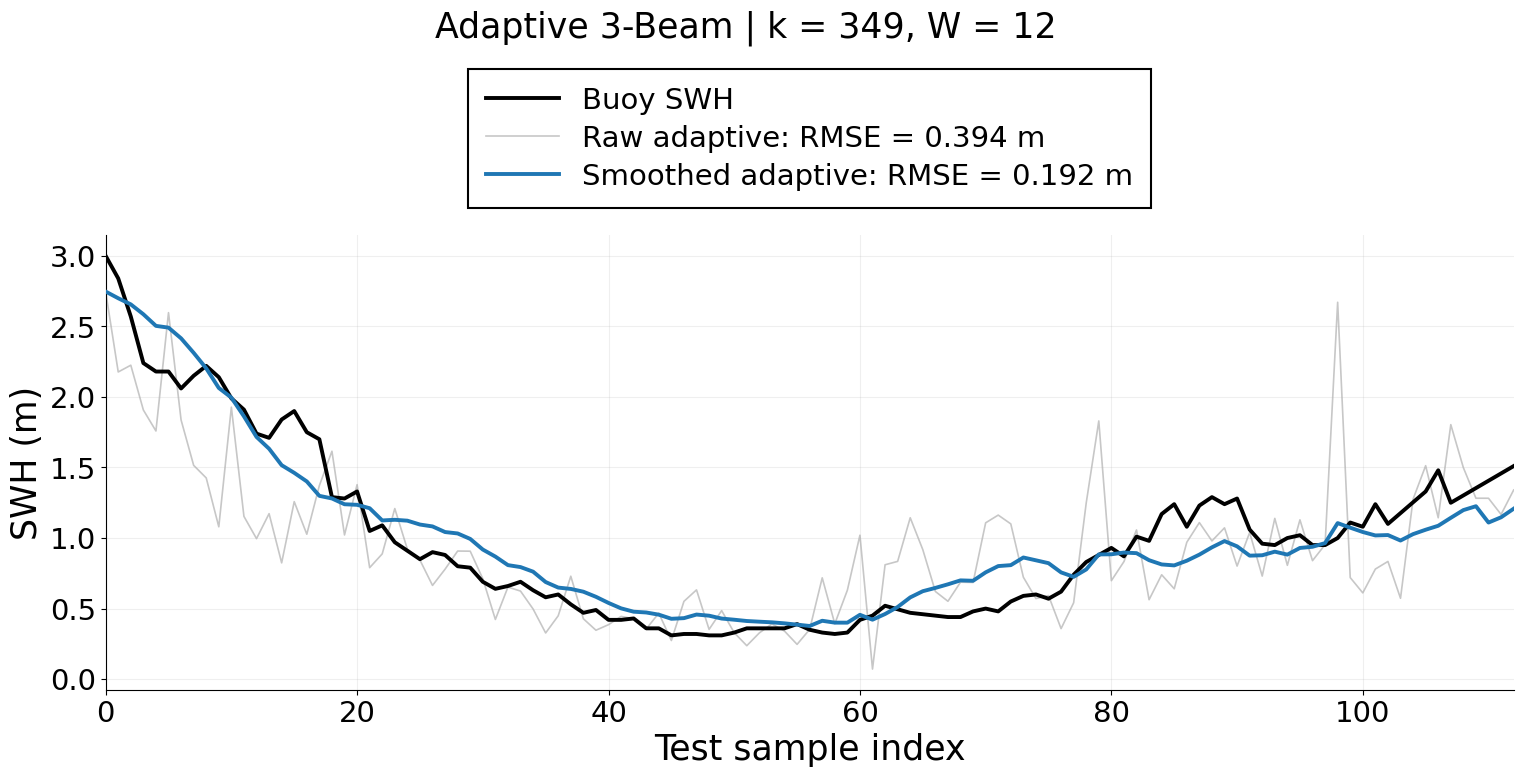}

    \caption{SWH predictions over the common
    evaluation interval for fixed Bartlett-based
    LOSE (top) and the adaptive three-beam
    estimator (bottom).}
    \label{fig:results_temporal_comparison}
\end{figure}

Fig.~\ref{fig:results_temporal_comparison}
shows the raw and smoothed predictions of
the fixed Bartlett-based LOSE and adaptive
three-beam estimators alongside the buoy
measurements. The buoy SWH initially declines
to approximately $0.3~\mathrm{m}$ before
recovering toward $1.5~\mathrm{m}$.
Both smoothed predictions reproduce this
broad evolution but underestimate the
initial SWH and several later maxima.

The raw Bartlett--LOSE predictions exhibit
pronounced sample-to-sample fluctuations,
including occasional negative estimates
and large positive excursions. The raw
adaptive predictions also fluctuate but
show smaller deviations overall, consistent
with their lower raw RMSE. The improvement
therefore occurs before temporal smoothing,
which attenuates the fluctuations in both
estimators.

The smoothed Bartlett--LOSE prediction
tracks the initial decline and the
low-wave-height interval but overestimates
portions of the early decline and
underestimates the subsequent rise.
During the later recovery, the buoy SWH
reaches approximately
$1.2$--$1.3~\mathrm{m}$ while the prediction
remains near $0.8$--$0.9~\mathrm{m}$.
At the end of the record, the prediction
reaches approximately $1.2~\mathrm{m}$,
compared with a buoy SWH of
$1.5~\mathrm{m}$.

The smoothed adaptive prediction similarly
captures the initial decline and
low-wave-height interval but overestimates
parts of the transition into the recovery
and underestimates several later maxima.
It also reaches approximately
$1.2~\mathrm{m}$ at the end of the record.
Thus, despite the lower raw and smoothed
RMSEs, discrepancies remain during changes
in sea state, and the reported smoothed
correlation does not increase.

\begin{figure}[!t]
    \centering
    \includegraphics[width=\linewidth]
    {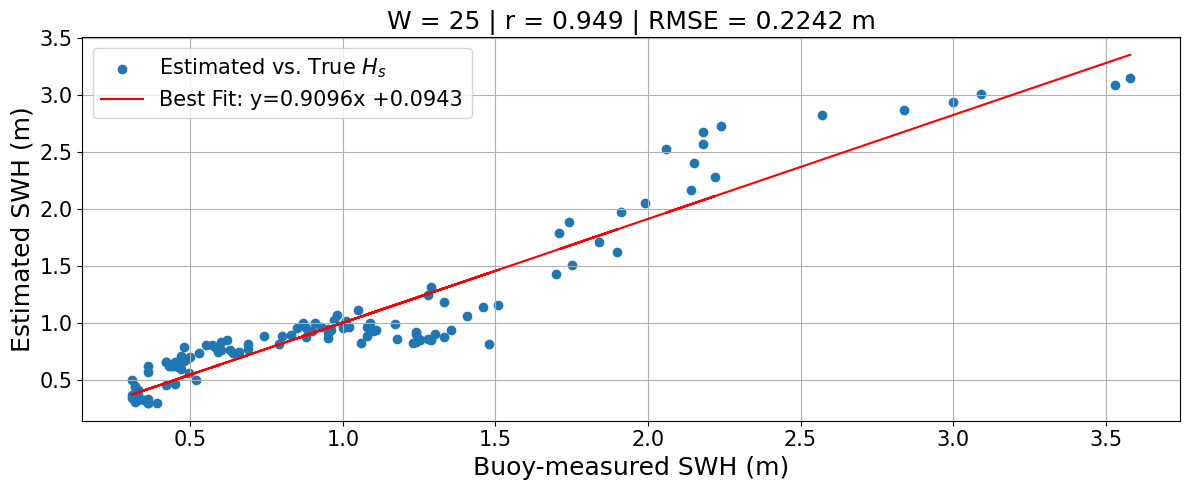}

    \medskip

    \includegraphics[width=\linewidth]
    {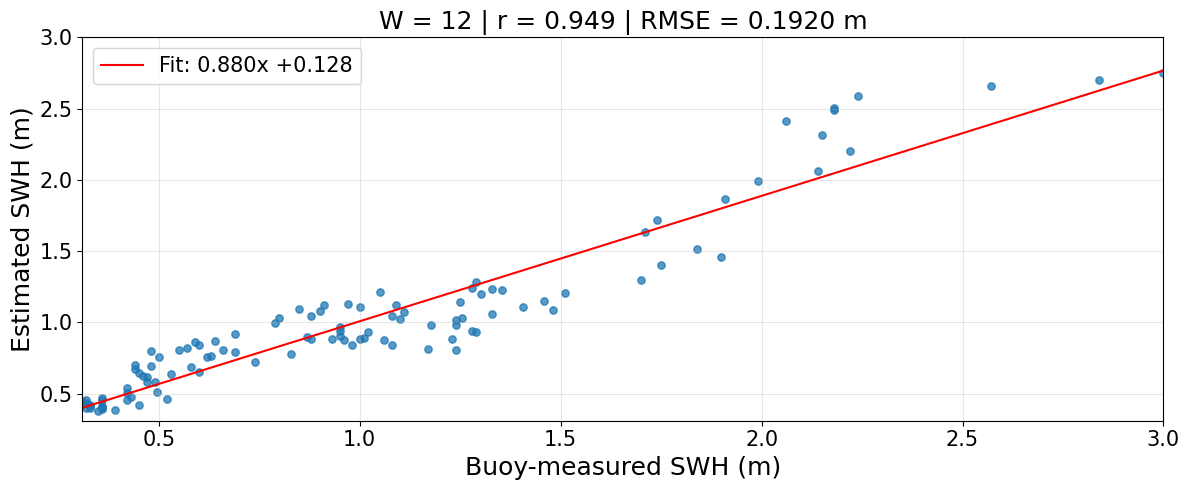}

    \caption{Smoothed estimated versus buoy-measured
    SWH on the common evaluation record for fixed
    Bartlett-based LOSE (top, $r=0.949$) and the
    adaptive three-beam estimator
    (bottom, $r=0.949$).}
    \label{fig:results_scatter_comparison}
\end{figure}

The scatter plots in
Fig.~\ref{fig:results_scatter_comparison}
complement the temporal comparison. For buoy
SWH values of approximately
$1.0$--$1.5~\mathrm{m}$, the adaptive estimates
are generally higher than the Bartlett-based
LOSE estimates, consistent with improved
tracking of the later recovery.
At the upper end of the observed range,
both estimators produce values near
$3.2~\mathrm{m}$ for buoy measurements
of approximately $3.5$--$3.6~\mathrm{m}$.
The adaptive configuration therefore provides
an overall reduction in estimation error,
although underestimation remains evident.

\subsection{Temporal Smoothing and Adaptive Prediction Stability}
\label{subsec:results_smoothing}

The selected adaptive configuration used
a causal averaging window of $W=12$
observation intervals. Its raw predictions
exhibit substantial sample-to-sample
fluctuations, including isolated negative
estimates and positive excursions exceeding
$3~\mathrm{m}$. These excursions are not
consistently associated with comparable
changes in the buoy record.

Causal sliding-window averaging attenuates
these fluctuations and produces the smoother
result associated with the reported
RMSE of $0.192~\mathrm{m}$. Temporal
postprocessing therefore contributes to
the stability of the adaptive output.
The same averaging mechanism is used in
the LOSE implementation, although the
selected window lengths need not be
identical. The reported accuracy comparison
is between the smoothed outputs.

\subsection{Comparative Assessment of Fixed and Adaptive
Spatial Processing}
\label{subsec:results_model_comparison}

Fixed Bartlett-based LOSE extracts ordered
magnitudes from a prescribed spatial output,
whereas the adaptive estimator combines
ordered magnitudes from three trainable
complex spatial outputs. For the selected
configurations, the adaptive pipeline reduces
smoothed RMSE from $0.224~\mathrm{m}$ to
$0.192~\mathrm{m}$, an improvement of $14.3\%$,
and maintained the correlation at $0.949$. These results show improved estimation performance
for the selected adaptive configuration on the
common evaluation record.

Fig.~\ref{fig:results_temporal_comparison}
shows similar performance during the initial
decline and minimum-SWH interval, with clearer
differences during the later recovery.
Around evaluation indices $80$--$90$,
adaptive estimates reach approximately
$0.8$--$0.9~\mathrm{m}$, compared with
$0.6$--$0.7~\mathrm{m}$ for fixed
Bartlett-based LOSE, while buoy SWH reaches
$1.2$--$1.3~\mathrm{m}$. Adaptation therefore
reduces underestimation during this recovery,
although it increases overestimation during
parts of the preceding low-SWH transition.
Consistently, Fig.~\ref{fig:results_scatter_comparison}
shows an upward shift of the low-estimate
cluster for buoy SWH between approximately
$1.0$ and $1.5~\mathrm{m}$. Both models,
however, underestimate the largest observed
wave heights and retain fitted slopes below
unity.

The improvement comes with additional
training requirements for the complex spatial
weights and a larger concatenated feature
representation. Fixed Bartlett-based LOSE
remains simpler and captures the main
sea-state evolution without spatial-weight
optimization. The adaptive raw predictions
also exhibit substantial excursions so the reported
accuracy advantage applies to the complete
pipeline with causal averaging.

\section{Conclusion and Future Work}
\label{sec:conclusion}
This study investigated an adaptive beamforming
approach to improve time-domain significant wave height
(SWH) estimation from multichannel HF radar
measurements. The motivation was to optimize
the beamforming weights for SWH estimation
accuracy rather than retain fixed weights
determined by array geometry and look direction.
The resulting extension of the linear
ordered-statistics estimator (LOSE) uses three
trainable complex spatial filters, ordered-magnitude
features, and least-squares regression.

On a common radar--buoy record with a chronological training--evaluation partition, the selected adaptive configuration reduced the sliding-window-smoothed RMSE from $0.224~\mathrm{m}$ to $0.192~\mathrm{m}$ ($14.3\%$), while Pearson correlation remained at $0.949$. Both estimators captured the broad SWH evolution but underestimated several maxima, and the raw adaptive predictions exhibited substantial short-term fluctuations. The improvement applies to the complete selected configurations, including their feature counts and temporal averaging.

Future work would benefit from evaluating longer records across additional sites and sea states, comparing fixed three-beam and adaptive single-beam configurations under matched conditions, and investigating regularization and the trade-off between smoothing and responsiveness.

\bibliographystyle{IEEEtran}
\bibliography{bibtex/bib/IEEEexample}

@phdthesis{Gill1999HFScattering,
  author       = {Eric W. Gill},
  title        = {The Scattering of High Frequency Electromagnetic Radiation from the Ocean Surface: An Analysis Based on a Bistatic Ground Wave Radar Configuration},
  school        = {Memorial University of Newfoundland},
  address       = {St. John's, Newfoundland, Canada},
  year          = {1999},
  type          = {Ph.D. Thesis},
  url           = {https://memorial.scholaris.ca/items/820779cc-74b3-40fd-bf57-3371ed4e4510}
}

@inproceedings{shahidi_chandrasekara_linear,
author = {Chandrasekara, Senal and Shahidi, Reza},
year = {2025},
month = {07},
pages = {2498-2501},
title = {An Accurate Physics-Based Single-Layer Neural Network for Significant Wave Estimation from High Frequency Radar Data},
doi = {10.1109/AP-S/CNC-USNC-URSI55537.2025.11266027}
}

@article{Barrick1977,
  author  = {Barrick, D. E.},
  title   = {Extraction of Wave Parameters from Measured {HF}
             Radar Sea-Echo Doppler Spectra},
  journal = {Radio Science},
  volume  = {12},
  number  = {3},
  pages   = {415--424},
  year    = {1977},
  doi     = {10.1029/RS012i003p00415}
}

@article{Wyatt1986,
  author  = {Wyatt, L. R.},
  title   = {The Measurement of the Ocean Wave Directional
             Spectrum from {HF} Radar Doppler Spectra},
  journal = {Radio Science},
  volume  = {21},
  number  = {3},
  pages   = {473--485},
  year    = {1986},
  doi     = {10.1029/RS021i003p00473}
}

@article{Hisaki1996,
  author  = {Hisaki, Yukiharu},
  title   = {Nonlinear Inversion of the Integral Equation to
             Estimate Ocean Wave Spectra from {HF} Radar},
  journal = {Radio Science},
  volume  = {31},
  number  = {1},
  pages   = {25--39},
  year    = {1996},
  doi     = {10.1029/95RS02439}
}

@article{WyattLimits2000,
  author  = {Wyatt, Lucy R.},
  title   = {Limits to the Inversion of {HF} Radar Backscatter
             for Ocean Wave Measurement},
  journal = {Journal of Atmospheric and Oceanic Technology},
  volume  = {17},
  number  = {12},
  pages   = {1651--1666},
  year    = {2000},
  doi     = {10.1175/1520-0426(2000)017<1651:LTTIOH>2.0.CO;2}
}

@article{WyattQuality2011,
  author  = {Wyatt, L. R. and Green, J. J. and Middleditch, A.},
  title   = {{HF} Radar Data Quality Requirements for Wave Measurement},
  journal = {Coastal Engineering},
  volume  = {58},
  number  = {4},
  pages   = {327--336},
  year    = {2011},
  doi     = {10.1016/j.coastaleng.2010.11.005}
}

@article{ShahidiInverse2021,
  author  = {Shahidi, Reza and Gill, Eric W.},
  title   = {A New Automatic Nonlinear Optimization-Based Method
             for Directional Ocean Wave Spectrum Extraction
             from Monostatic {HF}-Radar Data},
  journal = {IEEE Journal of Oceanic Engineering},
  volume  = {46},
  number  = {3},
  pages   = {900--918},
  year    = {2021},
  doi     = {10.1109/JOE.2020.3009770}
}

@inproceedings{Shahidi2018,
  author    = {Shahidi, Reza and Gill, Eric W.},
  title     = {Significant Wave Height Estimation without Doppler Spectra},
  booktitle = {2018 18th International Symposium on Antenna
               Technology and Applied Electromagnetics (ANTEM)},
  pages     = {1--2},
  year      = {2018},
  doi       = {10.1109/ANTEM.2018.8572884}
}

@article{Shahidi2020,
  author  = {Shahidi, Reza and Gill, Eric W.},
  title   = {Two New Methods for the Extraction of Significant
             Wave Heights from Received {HF}-Radar Time Series},
  journal = {IEEE Geoscience and Remote Sensing Letters},
  volume  = {17},
  number  = {12},
  pages   = {2070--2074},
  year    = {2020},
  doi     = {10.1109/LGRS.2019.2961917}
}

@article{Hashemi2023,
  author  = {Hashemi, Sepideh and Shahidi, Reza and Gill, Eric W.},
  title   = {Extraction of the Significant Wave Height from Synthetic
             {HF} Radar Data Acquired on a Floating Platform},
  journal = {IET Radar, Sonar \& Navigation},
  volume  = {17},
  number  = {11},
  pages   = {1639--1645},
  year    = {2023},
  doi     = {10.1049/rsn2.12452}
}

@article{MoralesMarquez2025,
  author  = {Morales-M{\'a}rquez, Ver{\'o}nica and
             Dumas, Dylan and Gu{\'e}rin, Charles-Antoine},
  title   = {{HF} Radar Estimation of Ocean Wave Parameters:
             Second-Order Doppler Spectrum versus Bragg Wave
             Modulation Approach},
  journal = {Coastal Engineering},
  volume  = {198},
  pages   = {104719},
  year    = {2025},
  doi     = {10.1016/j.coastaleng.2025.104719}
}

@misc{ChandrasekaraWESE2026,
  author       = {Chandrasekara, Senal and Shahidi, Reza},
  title        = {Windowed Envelope Statistics for Time-Domain
                  Significant Wave Height Estimation from {HF} Radar},
  year         = {2026},
  howpublished = {arXiv preprint arXiv:2609.10920},
  url          = {https://arxiv.org/abs/2609.10920}
}

@article{Penrose1955,
  author  = {Penrose, Roger},
  title   = {A generalized inverse for matrices},
  journal = {Proceedings of the Cambridge Philosophical Society},
  year    = {1955},
  volume  = {51},
  number  = {3},
  pages   = {406--413},
  doi     = {10.1017/S0305004100030401}
}

@misc{ChandrasekaraQSOCE2026,
  author       = {Chandrasekara, Senal and Shahidi, Reza},
  title        = {Significant Wave Height Estimation Incorporating
                  Second-Order Scattering},
  year         = {2026},
  howpublished = {arXiv preprint arXiv:2603.17116},
  url          = {https://arxiv.org/abs/2603.17116}
}

@article{VanVeen1988,
  author  = {Van Veen, Barry D. and Buckley, Kevin M.},
  title   = {Beamforming: A Versatile Approach to Spatial Filtering},
  journal = {IEEE ASSP Magazine},
  volume  = {5},
  number  = {2},
  pages   = {4--24},
  year    = {1988},
  doi     = {10.1109/53.665}
}

@article{Vorobyov2003,
  author  = {Vorobyov, Sergiy A. and Gershman, Alex B. and
             Luo, Zhi-Quan},
  title   = {Robust Adaptive Beamforming Using Worst-Case Performance
             Optimization: A Solution to the Signal Mismatch Problem},
  journal = {IEEE Transactions on Signal Processing},
  volume  = {51},
  number  = {2},
  pages   = {313--324},
  year    = {2003},
  doi     = {10.1109/TSP.2002.806865}
}
\vfill
\end{document}